\documentclass[12pt,a4paper]{article}
\usepackage[a4paper, total={6in,9in}]{geometry}

\usepackage{amsmath}
\usepackage{amsfonts}
\usepackage{amssymb}
\usepackage{graphicx}
\usepackage{color}
\usepackage{epstopdf}
\usepackage{xcolor,colortbl}
\usepackage[thinlines]{easytable}
\usepackage{caption}
\usepackage{placeins}

\newcommand{\LB}[1]{{}}

\newcommand{\psiII}[0]{\psi_\mathrm{TF}}

 \usepackage{ulem}

\title{Capillary freezing of ionic liquids confined at metallic interfaces}
\title{Electronic effects on the confinement induced capillary freezing of ionic liquids}
\title{Capillary freezing of ionic liquids confined between metallic interfaces}
\title{Electronic effects on the capillary freezing of ionic liquids confined between metallic interfaces}
\title{Electronic screening effects on the capillary freezing of ionic liquids confined between metallic interfaces}
\title{Nanoscale capillary freezing of ionic liquids confined between metallic interfaces and the role of electronic screening}
\author
{\normalsize{Jean Comtet,$^{1}$ Antoine Nigu\`{e}s,$^{1}$ VojteÂch Kaiser,$^{1}$} \\
\normalsize{Lyd\'{e}ric Bocquet,$^{1}$ Alessandro Siria,$^{1}$}\\
\small{$^{1}$Laboratoire de Physique Statistique, Ecole Normale Sup\'{e}rieure, }\\
\small{UMR CNRS 8550, 
PSL Research University,}\\
\small{75005 Paris Cedex 05, France}\\
}

\begin{document}

\baselineskip24pt

\maketitle
\begin{abstract}
\linespread{1.6}

\normalsize{\bf
Room temperature Ionic liquids (RTIL) 
received considerable attention as a new class of materials with fundamental importance for energy storage \cite{Armand2009, Uesugi2013} and active lubrication \cite{Palacio2010, Dold2015, Smith2013a}. 
Their unique properties result from the competition of strong electrostatic interactions with properly designed molecular structure to avoid crystalization at room temperature. They are however unsual liquids, which challenge fundamentally the classical frameworks of electrolytes. In particular their behavior at electrified interfaces remains elusive with very rich and exotic responses relevant to their electrochemical activity \cite{Kornyshev2014,Merlet2012,Perkin2013,Atkin2009,Rotenberg2015,Endres2012,
Bovio2009a,Bovio2009,Yokota2010}.
In this work, we use quartz tuning fork based AFM nanorheological measurements to explore the properties of RTIL in nanometric confinement.
We unveil a dramatic change of the RTIL towards a solid-like phase below a threshold thickness, pointing to a capillary freezing in confinement. This threshold thickness is found to be intimately related to the metallic nature of the confining materials, with more metallic surfaces facilitating capillary freezing. This behavior is interpreted theoretically in terms of the shift of the freezing transition, taking into account the influence of the electronic screening on RTIL wetting of the confining surfaces, as described by the simple Thomas-Fermi approach.
Our findings provides fresh views on the properties of confined RTIL with important implications for their statics and dynamics inside nanoporous metallic structures. {This also suggests applications to tune nanoscale lubrication with the phase-changing RTIL, by varying the nature as well as the patterning of the subtrate, and the application of active polarisation.}
}
\end{abstract}

\maketitle

%
%
%

The confinement of liquids at nanoscales leads to a broad spectrum of new properties which are harnessed in a variety of applications, 
from energy storage
, friction to catalysis \cite{Siria2013}. The nanoscale realm hosts indeed a broad spectrum of molecular forces that compete to make new fluid behavior emerge \cite{Bocquet2010,Secchi2016,Secchi2016a}. 
Reversly, nanoscale confinement is a fine probe which allows to disentangle the molecular mechanisms at play. 
In this work, we explore the mechanical bevavior of Room Temperature Ionic Liquid (RTIL) in nanoscale confinement. Such system is  a prototype for a dense electrolyte -- composed here of pure ions -- and accordingly, electrostatic interactions do control the behavior of these liquids. But at such densities, standard mean-field response, which constitute the toolbox of dilute electrolytes, cannot account for
the structure of the electric double layer close to (charged) surfaces. Confinement therefore opens an interesting window on the physics of dense electrolytes and their interaction with the confining interfaces. In particular, due to the dominant role of electrostatic forces, one may anticipate that the metallic nature of the confining surfaces should affect the static and dynamic properties of confined RTIL.
Such relationship has not been explored up to now.

\begin{figure}[!htb]
\centering
    \includegraphics[width=\columnwidth]{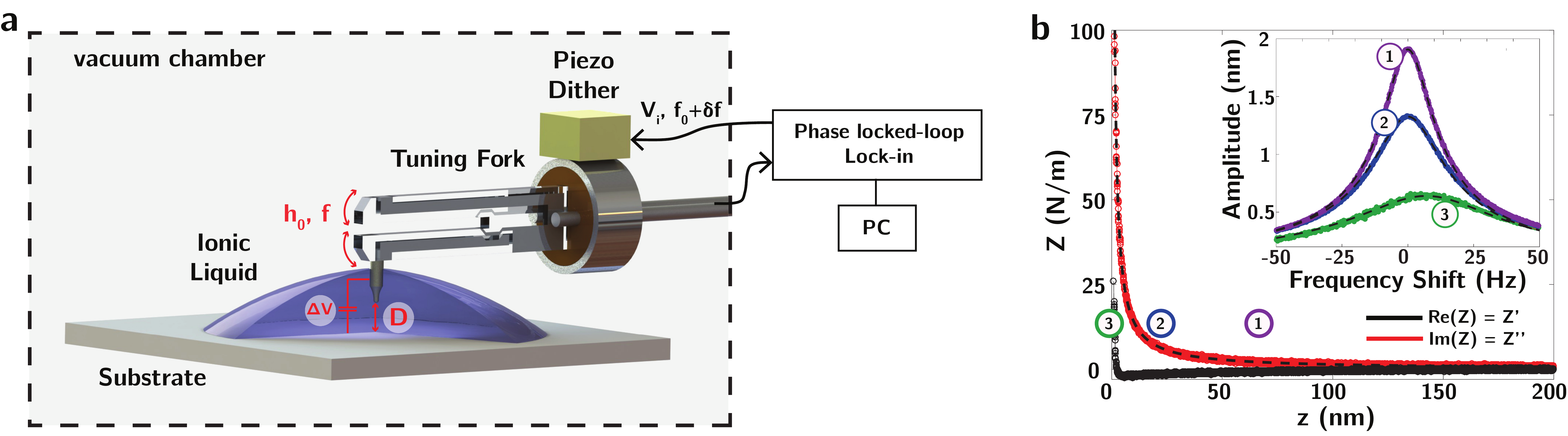}
      \caption{\label{etiquette} \textbf{Experimental Set-up.} \textbf{(a)} Schematic of the experiment. An etched tungsten tip of end radius of curvature R between 50 nm and $2.5~\mu$m is glued to the tuning fork, and immersed in the Ionic Liquid. The tuning fork is excited by a piezo dither at a frequency $f_0\approx 32$~kHz, and a lock-in and a phase-locked loop maintain constant both the oscillation amplitude $h_0$ and the phase shift between the tuning fork and the excitator. 
      The distance $D$ between the tip and the substrate is controlled through a piezo-element with sub-nanometric resolution in displacement.
    The substrate can be biased with respect to the tip of a potential difference $\Delta V$. The experimental set-up is placed in a vacuum chamber at a pressure of $\approx10^{-6}$ mbar. \textbf{(b)} Nanorheological measurement on a silicon oil confined between a tungsten tip and a mica sample, showing the variation of conservative (Z' - black) and dissipative (Z'' - red) mechanical impedance (see Eq. (1)); black dotted line is a fit based on Eq.~(2). Inset shows typical resonance curves of the tuning fork with the tip immersed in the liquid and far from the surface (1), close to the surface (2) and in contact with the substrate (3).}
\end{figure}

{\it Experimental Set-up --} We present in Fig.~1a a sketch of the experimental set-up, see Supplementary Material section 1 for more details. Briefly, we glue an electrochemically etched tungsten tip of end radius from 50 nm to 2.5 $\mu$m to a milimetric quartz tuning fork, which serves as our force sensor.
The tuning fork is excited via a piezo-dither, and the oscillation amplitude and phase shift of the tuning fork with regards to the excitation voltage are monitored through the piezoelectric current flowing through the tuning fork electrodes. By mean of a piezo-element with sub-nanometric resolution in displacement, the probed liquid is confined between the oscillating tunsten tip and substrates of various nature. Importantly, the entire set-up is placed in a vacuum chamber at a pressure of approximately $10^{-6}$ mbar. 

This set-up allows us to measure the mechanical impedance $Z^*= F^*/h_0 = Z'+i\cdot Z''$ of the confined liquid, defined as the ratio of the complex amplitude of the dynamic force $F^*$ acting on the tip, to the amplitude $h_0$ of the tip oscillation. The inset of Fig. 1b shows a typical resonance curve of the tuning fork for a fixed excitation voltage of the piezoelectric element, with the tip immersed in a newtonian silicon oil (1) far from the surface, (2) close to the surface and (3) in contact with the substrate. As the interaction of the tip with its environment is modified, one observes a change in both the resonance frequency and the amplitude at resonance. The shift in resonance frequency $\delta f$ is related to the conservative force response Z', whereas the broadening of the resonance (change of quality factor $Q_0 \rightarrow Q_1$) is related to dissipation Z'' \cite{Nigues2014}. During a typical experiment, two feedback loops allow us to work at the resonance and  maintain constant the oscillation amplitude $h_0$ of the tuning fork. Monitoring the frequency shift $\delta f$ and the excitation voltage $V_i$ thus provides a direct measurement of real ($Z' = \text{Re}(Z^*)$) and imaginary ($Z''= \text{Im}(Z^*)$) part of the mechanical impedance:
%

The experimental set-up has been fully benchmarked using a newtonian silicon oil, as shown in Fig. 1b.
%
%
\begin{eqnarray}
Z' = 2 K_0 \frac{\delta f}{f_0} \text{ and } Z'' = \frac{K_0}{\sqrt{3}} \left( \frac{1}{Q_0} -\frac{1}{Q_1} \right)
\end{eqnarray}
where $f_0$ is the bare resonance frequency and $K_0$ [N/m] is the tuning force spring constant. The advantages of the tuning fork are twofold: first, an ultrahigh stiffness of $K_0\approx~40$~kN/m which prevents any mechanical instability during the approach and second, very low oscillation amplitude (0.1 - 2 nm) together with very low intrinsic dissipation characterized by high quality factors of up to tens of thousands in vacuum and in the range of few thousands even when the tip is immersed in high viscosity liquid. These characteristics made the tuning fork AFM the ideal instrument to study, for example, tribology in individual nanostructures such as nanowires and nanotubes \cite{Nigues2014}.

The ionic liquid under investigation is BmimBF4 
(Sigma Aldrich, 98.5\% purity), which is further filtered through a 100~nm hole teflon membrane before use. A drop is deposited on the substrate and the AFM tungsten tip is immerged in the liquid. 
The liquid is left at rest in the vacuum chamber at least for 12h to remove  water impurities. The substrate can be biased with respect to the tip by a potential difference $\Delta V$. To verify the high purity of the ionic liquid,  we systematically check the absence of long-term electrochemical current when applying a potential drop  between -1.8 V and 1.8 V, which is smaller than the electrochemical window for this liquid \cite{OMahony2008}.

We have explored various substrates, namely Mica, HOPG, doped silicon, and platinum, whose characteristics are described in the Supplementary Materials section 2. Note also that platinum and doped silicon may be coated by natural oxide layers of up to 1 nm in thickness \cite{Morita1990, Anson1957}, but we anticipate from our results below that this value is much smaller than the typical length at which the phenomena under investigation occur, in the range of tens of nanometers.

\begin{figure}[!htb]
\centering
    \includegraphics[width=0.8\columnwidth]{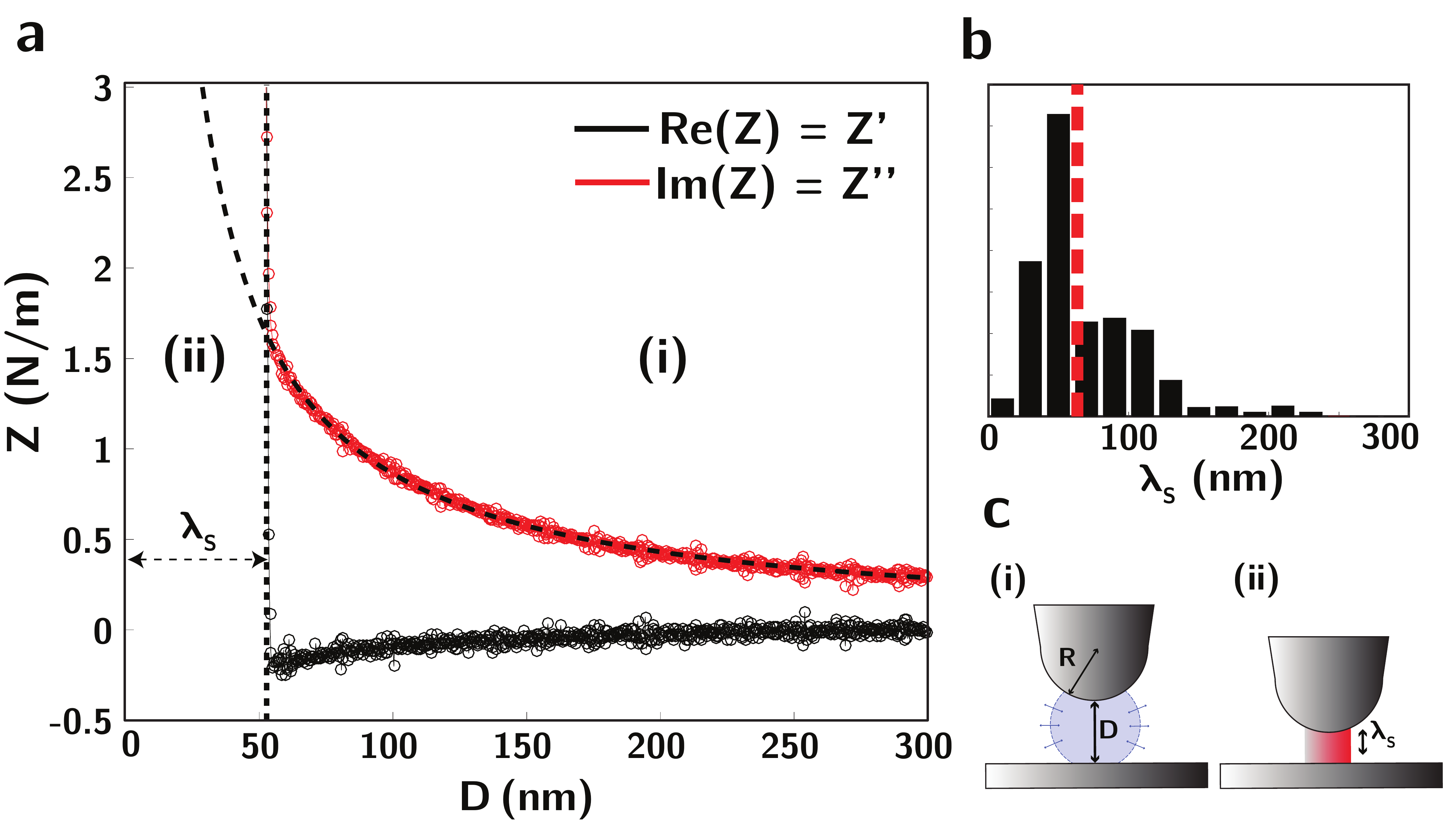}
      \caption{\textbf{ Confinement induced freezing transition:} {\bf a)} Real ($Z'$, black) and complex  ($Z''$, red) part of the mechanical impedance characterizing conservative and dissipative response of the liquid, versus tip-substrate position $D$, performed on a HOPG sample. 
      The origin of the tip-substrate position is defined as the hydrodynamic zero obtained by fitting $Z''$ with equation (2). Upon a critical confinement distance $\lambda_\text{S}$, $Z^*$ changes from a liquid-like response to a solid-like response, characterized by the onset of an elastic contribution ($Z'$, black dots) and a sharp increase of the dissipation ($Z''$, red dots). {\bf b)} Distribution of $\lambda_\text{S}$ on HOPG; $\approx$ 12,000 cycles were performed, over 18 distinct locations. Red vertical line is the mean of the distribution. {\bf c)} Sketch of the freezing induced by confinement: upon a critical confinement the RTIL changes from liquid (i) to solid-like~(ii). }
      \label{ZZ}
\end{figure}

{\it Confinement induced freezing --} We plot in Fig.~\ref{ZZ}a the typical variation of the elastic  $Z'=\text{Re}(Z^*)$ (black curve) and dissipative $Z''=\text{Im}(Z^*)$ (red curve) parts of the mechanical impedance, as the tungsten tip approaches a HOPG surface in the ionic liquid. Far from the substrate - zone (i) - the elastic response $Z'\approx 0$ within the experimental precision; one can observe a minute attractive component reminiscent of what is observed in \cite{perkins}. The dissipative component $Z''$ increases gently as the confinement thickness decreases; quantitatively, the increase of $Z''$ with decreasing confinement $D$ can be described by the Reynolds dissipative response of a viscous newtonian fluid sheared under an oscillating sphere \cite{Leroy2011}:
\begin{eqnarray}
Z^*(D) = i\frac{6\pi \eta  R^2  \cdot 2\pi f_0}{D}
\label{hydro}
\end{eqnarray}
where $R\approx 1-2.5~\mu$m is the radius of curvature of the tip, $\eta \approx 0.15$ Pa.s the liquid bulk viscosity, $f_0~\approx~32$~kHz the oscillation frequency and $D$ [nm] the distance between the tip and the substrate. This expression assumes no slip at the liquid/substrate interfaces, which is expected for such strongly interacting systems \cite{Bocquet2010}. To avoid errors induced by the real geometry of the tungsten tip, see Fig S1 in Supplementary Materials, we calibrated the technique on a silicon oil with viscosity $\eta \approx 0.1$ Pa.s, comparable to $\eta \approx 0.15$ Pa.s for ionic liquid; As shown on Figs. 1c and 2a, this prediction reproduces very well the experimental data and allows to determine the ``hydrodynamic zero'', which defines the absolute distance $D$ between the tip and the substrate, see Supplementary Materials Section 1.


As shown in Fig.~2a, in ionic liquids, before reaching the hydrodynamic "zero" $D\rightarrow 0$, both the elastic $Z'$ and dissipative part $Z''$ of the response suddenly diverge at a critical confinement $D=\lambda_S$. This occurs for a confinement $D$ in the range of a few tens of nanometers, depending on the substrate. 
We have changed the tip oscillation amplitude $h_0$ over one decade, between 0.1 nm and 1 nm, verifying that $\lambda_S$ does not depend on oscillation amplitude and shear rates.

This strong repulsive elastic reponse (with $Z' \approx 30$~N/m) shows that the ionic liquid can now sustain a yield stress of order $\tau~\approx~Z' h_0/\pi R^2 \approx$ 1 kPa, providing a clear signature of the solid-like response of the confined RTIL for $D<\lambda_S$.
This behavior was found repeatedly when performing approach and retract cycles of the tip, either at the same or at distinct locations;
typically $\approx$ 10000 cycles were performed for each material, over $\approx$~20 distinct locations.
We measured accordingly the distribution of threshold confinement thickness, as reported in the inset of Fig. 2~b for HOPG, allowing to extract the mean transition thickness, found to be $\lambda_S \simeq$ 60 nm for HOPG.

Going further, the same phenomenon was observed for the various substrates under investigation, Mica, HOPG,  doped Silicon and Platinum,
with a mean transition thickness $\lambda_S$ increasing in the order: Mica (15 nm) $<$ HOPG (60 nm) $<$ doped Silicon (110 nm) $<$ Platinum (160 nm). Interestingly this order corresponds to substrates with increasing metallic character, as for example characterized by the conductivity, from the insulating Mica to the highly conductive Platinum.

Finally, we have explored the influence of a voltage drop $\Delta V$ across the RTIL on this transition and found that $\lambda_S$ varies by typically 20 nm, when changing the voltage drop between the two substrates, see Supplementary Materials Fig. S4. 

{\it Prewetting, capillary freezing and electronic screening --}
As a first interpretation of these results, one may infer the presence of solid layers pre-existing on the surface of the substrates. 
To explore this assumption, we have performed AFM images of the surfaces using a sharp tip with a 10 - 50 nm of
radius of curvature, see Supplementary Materials Fig. S1. While in vacuum, the substrates surface appear atomically smooth on micrometric scales -- with a typical rms roughness between 0.3 nm and 1 nm depending on the substrate, see Supplementary Materials Fig. S2 --
one indeed observes  solid-like terrace structures on the surfaces when immersed in RTIL. This thickness is measured below 1 nm for HOPG  and typically in the range of $\sim 20-30$ nm for doped silicon and platinum; no such terrace is evident on Mica, see Supplementary Materials Fig. S3. Such structures are reminiscent of observations using STM and AFM imaging \cite{Endres2012,Yokota2010,Elbourne2015,Buchner2016}. 
That such thick structures are present on the substrate surfaces is unexpected {\it per se} and raises the question of the prewetting of the surfaces by the RTIL, and the role of the metallic nature of the substrates on this prewetting. We note however that the characteristic height of these solid ``prewetting'' films is much smaller than the critical thickness at which the transition occurs for each substrates, see Fig.~3, blue dots. Accordingly an alternative thermodynamic explanation should be sought.


 \begin{figure}[!htb]
\centering
   \includegraphics[width=0.7\columnwidth]{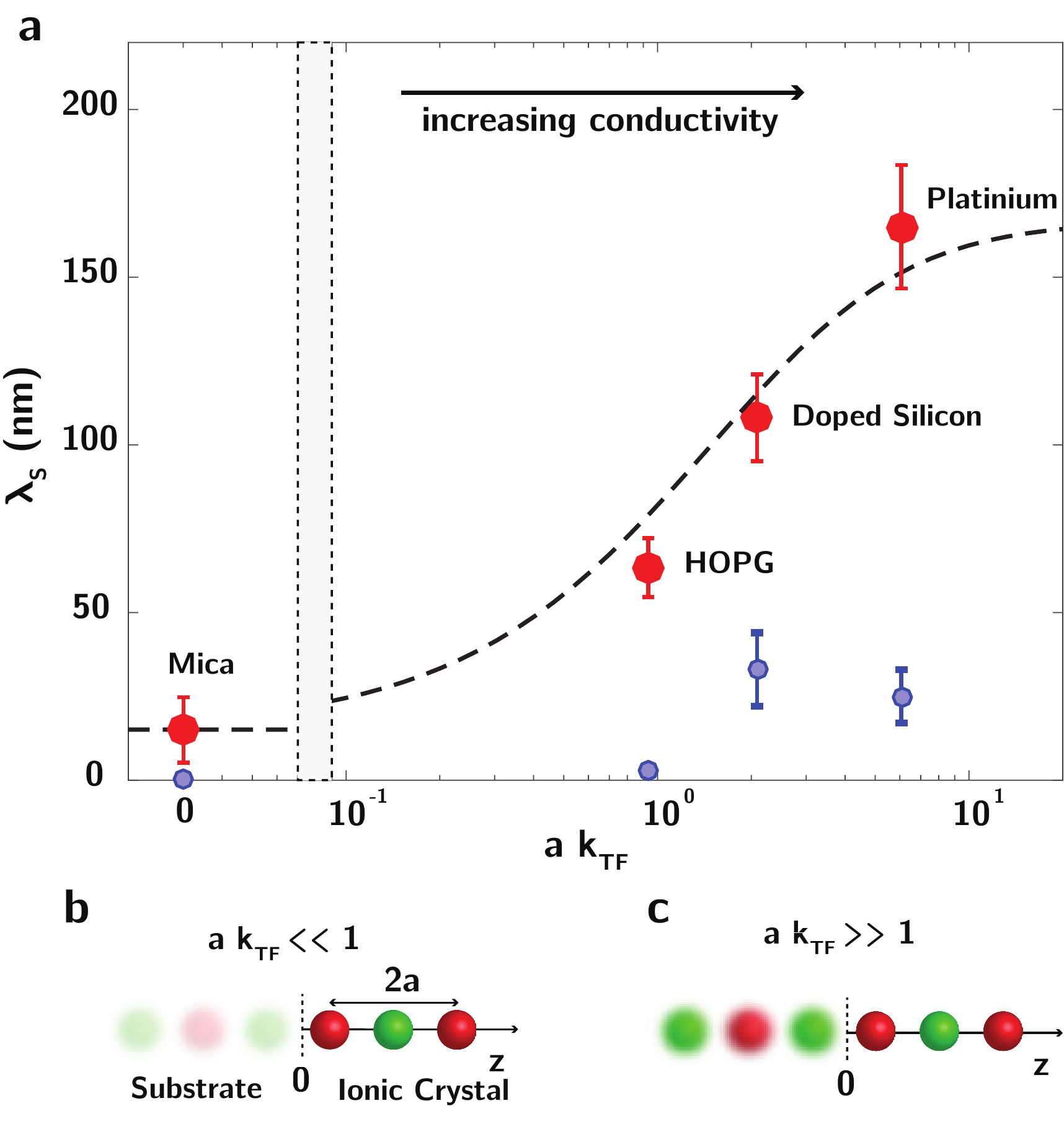}
      \caption{\textbf{Effect of substrate electronic properties on the freezing transition. a)} Red dots: Variation of the mean solidification length (red dots) $\lambda_\text{S}$ on Mica and three conductive substrates: HOPG, Doped Silicon and Platinum versus the normalized Thomas-Fermi wavevector, $a\cdot k_\text{TF}$, wih $2a$ the typical ionic crystal lattice constant. Error bars indicate standard deviation for Mica and standard error for the three conductive substrates, with $N\approx 20 $ the number of positions investigated in each substrates. Blue open dots represent the typical size of the preexisting solid layers measured with a sharp tip. Error bar represents standard deviation of estimated height. Dotted line is the prediction from Eq. (7).
        \textbf{b-c)} Schematic representation of the ionic crystal close to an insulator ($ \alpha k_\text{TF}  \ll 1$, {(b)}) and a perfect metal ($ \alpha k_\text{TF}  \gg 1$, {(c)}) for which image charges induced in the substrate decrease the energy of the system.}
        \label{TF}
        \end{figure}

Looking at  Fig. 2a, the drastic change in both elastic response and dissipation from $D>\lambda_S$ to $D<\lambda_S$ would actually rather suggest a complete confinement induced phase change, with a freezing of the confined RTIL inside the gap.
A confinement induced phase transition is expected when the defavorable bulk free energy is balanced by favorable (wetting) surface contributions, thus stabilizing the unfavored phase in the confinement. This results in a shift for the phase transition, as observed for capillary condensation (the shifted liquid-gas phase transition) or the capillary freezing (shifted crystalization) \cite{Alba-Simionesco2006}. The balance of free energy leads to the so-called Gibbs-Thompson equation,
which characterizes the critical confinement $\lambda_S$ at which the free energies of the liquid and solid phase become equal \cite{Alba-Simionesco2006}: 
\begin{eqnarray}
\label{GT}
\Delta T = T_C - T_B = 2 \frac{T_B \Delta \gamma}{\rho\, L_h\,\lambda_S } \text{ with } \Delta \gamma = \gamma_{wl}-\gamma_{ws} 
\end{eqnarray}
where $\Delta T=T_C-T_B$ is the shift in transition temperature, in confinement $T_C$ as compared to the bulk transition occuring at $T_B$.
For the specific ionic liquid used here, $T_B=-71^\circ$~C; $\gamma_{wl}$ and $\gamma_{ws}$ are the surface energy of the liquid and solid phase with respect to the wall/substrate, $\rho=1.21$~g/mL the density of the liquid phase and $L_h= 47$ kJ/kg the latent heat of melting \cite{Bhatt2010} (see supplementary Table S1). Eq. (\ref{GT}) shows that if wetting of the solid-phase on the substrate is favored compared to that of the liquid ($\gamma_{ws}<\gamma_{wl}$) the freezing temperature of the confined phase $T_C$ is larger than the bulk freezing temperature $T_B$ ($T_C > T_B$ in Eq. (\ref{GT})). Putting numbers, one gets $T_C\sim 25^\circ$~C for a RTIL confined in a gap of $\sim 20$ nm with $\Delta \gamma \sim 0.3$ J/m$^2$ (anticipating on the values below). In other words, the RTIL may freeze in nanoconfinement at room temperature.
In this scenario, the distribution of confinement length measured experimentally, see inset if Fig.~3, can be understood as a signature of activation due to the first order character of the freezing transition (potentially facilitated by the prewetting phase on the substrate).


A delicate question though is to understand the variations with the metallic nature of the substrate. Following the argument above in terms of the shifted liquid-solid transition, this raises the question of the crystal-substrate surface energy and how it is influenced by the metallic character of the substrate. 
Physically, one may propose a simple explanation in terms of image charges. To highlight the argument, let us consider a semi-infinite ionic crystal at the interface with a perfect metal,
as sketch in Fig.~3b. The network of image charges builds a crystal structure with a (nearly) perfect symmetry with respect to the real upper half-lattice. Accordingly one expects the electrostatic contribution to the surface free energy to (nearly) vanish, as the system behaves as a single bulk lattice: $\gamma_{ws}^{elec}\rightarrow 0$. This requires of course a perfectly symmetric cristaline structure and this cancellation is not expected to occur for insulating substrates, or for disordered liquid phases. In other words, the (semi-infinite) ionic crystal has a lower surface energy at the interface with a metal wall as compared to an insulating substrate: $\gamma_{ws}^{insulating} > \gamma_{ws}^{metal}$.
This shows that the crystal phase is favored on metallic surface as compared to an insulating one  and the Gibbs-Thompson equation (Eq.~3) accordingly implies that the threshold confinement for the freezing transition should be larger with metal  as compared to insulating confining
surfaces.


In order to account for the non-ideal nature of  the metallic nature of the confining walls, one should model the electronic screening inside the substrates. To this end, we use the simple Thomas-Fermi framework, based on a local density approximation for the free electrons gas \cite{mahan2000}. 
This description provides a simple screening equation for the electric potential $V$ (V) in the metal, 
where the screening length 
$\lambda_\text{TF} = 1/k_\text{TF}$ [m] characterizes the typical length over which a defect charge is screened in the metal and is defined in terms of the density of states at the Fermi level, according to $k_\text{TF}^2= 4\pi e^2 {\partial n\over \partial \epsilon_F}$; $n_T$ the state occupation and $\epsilon_F$ the Fermi level. The limit of large $k_\text{TF}$ (small $\lambda_\text{TF}$) corresponds to the perfect metallic case for which $V$ is uniformly zero.
The electronic screening therefore modifies the interactions of charge close to the liquid-wall interface and the Green function $\Psi$ for the electrostatic interaction,
replacing the Coulomb interaction, obeys  equations:
\begin{eqnarray}
&\Delta \psi  &= -\frac{Q}{\epsilon }\delta(\mathbf{r}-\mathbf{r_0}) \,\,\,\text{for}\quad z>0\; \nonumber\\
&\Delta \psi &- k_\text{TF}^2\psiII = 0 \,\,\,\text{for}\quad z<0,
\end{eqnarray}
This allows to calculate the energy of the crystal in contact with the metal wall. Indeed 
the energy of this  {\it semi-infinite} crystal can be written in terms of the Green function and  charge density $\rho_\mathrm{cr}(\mathbf{r})= Q\sum_{n}(-1)^n\delta(\mathbf{r}-\mathbf{R_n})$, with $R_n$ the lattice sites, as $U = \frac{1}{2}\int  \mathbf{dr}   \rho(\mathbf{r})\psi(\mathbf{r})$. 
To proceed analytically, we consider the simplified situation of a 1D crystal which is expected to capture the main ingredients at stake.
The calculation of the energy is detailed in the Supplementary Materials, Section 5. The excess energy as compared to ideal metal limite
($k_\text{TF}\rightarrow \infty$) is found to take generically the form
\begin{equation}
\Delta U  =  \frac{Q^2}{16\pi\epsilon_0a} \mathcal{F}(k_\text{TF}a)
\end{equation}
with $a$ the lattice constant; the full expression of the function $\mathcal{F}(k_\text{TF}a)$ is given in Supplementary Eq. (11).
In line with the Laplace estimate of the surface tension \cite{RW}, we approximate the electrostatic contribution to the substrate-crystal surface tension as $\gamma_{ws}^{el} \simeq \Delta U$.  Furthermore one expects that the electrostatic contribution to the liquid-substrate surface tension to be substantially smaller than $\Delta U$.
Indeed 
the averaged charge density in the liquid is vanishing except very close to the wall due to the very strong screening in the RTIL, so that the corresponding electrostatic energy is expected to be smaller than the substrate-crystal contribution.
One can therefore neglect this contribution to the surface free energy as compared to the crystal surface tension and
write the electrostatic contribution to the excess surface energy as $\Delta \gamma^{el}= \gamma_{wl}^{el}-\gamma_{ws}^{el} \simeq \Delta U$.

As detailed in the Supplementary Materials Sec. 5, this simplifying assumption allows to express the difference $\Delta \gamma= \gamma_{wl}-\gamma_{ws}$ in surface energies as:
\begin{eqnarray}
\Delta \gamma=\Delta \gamma_0 \left(1+ \delta\cdot \mathcal{F}(k_\text{TF}a)\right) \text{ with } \delta =\frac{e^2}{16 \pi \epsilon_0 a^3 \Delta \gamma_0}
\label{DG}
\end{eqnarray}
where $\Delta \gamma_0$ [J/m$^2$] represents the non-electrostatic contributions to the surface energy (van der Waals, ...) and the possible (constant) contribution from the tungsten tip; $k_\text{TF}$ [m$^{-1}$] is the Thomas-Fermi wavevector as described above, $2a$ is the ionic crystal lattice constant, $e$ [C] is the elementary ionic charge, $\epsilon_0$ [F.m$^{-1}$] the ionic crystal dielectric constant; $\mathcal{F}\sim\mathcal{O}(1)$ is an increasing function whose expression is given and derived in the Supplementary Materials Sec. 5. 
The parameter $\delta$ compares the electrostatic contribution to surface energies, $\gamma^{elec} \sim {e^2}/{16 \pi \epsilon_0 a^3}$ to their non-electrostatic counterparts, $\sim \Delta\gamma_0$ and one expects $\delta=\gamma^{elec}/\Delta \gamma_0 \gg 1$.

Now using the Gibbs-Thompson result, Eq.~3, one predicts that the increase in surface energy difference for better metals, i.e. for larger Thomas-Fermi wavevector $k_{TF}$ (Eq.~\ref{DG}) will lead to a shift in the critical confinement distance for the freezing transition according to 
\begin{equation}
\lambda_S = \lambda_S^0(1+\delta \cdot \mathcal{F}(k_\text{TF}a)). 
\end{equation}
with $\lambda_S^0$ the value for the perfectly insulating material defined in terms of $\Delta \gamma_0$ (here Mica).

In Fig.~3, we compare the prediction for $\lambda_S$ with the experimental data for the various substrates investigated.
Note that in doing this comparison, we estimated the values of Thomas-Fermi length based on the substrate conductivity and carrier density, see Supplementary Material Sec. 2. We also fixed the lattice constant to 
$2a = 0.67$ nm (as estimated from the molar volume of the IL). 
As shown in Fig.~3, a good agreement between the theoretical predictions and the experimental results is obtained, yielding 
$\lambda_S^0=15$ nm and $\delta=10.1$. From the value for $\lambda_0$ and Eq. 3, one gets $\Delta \gamma_0 \approx 0.2$ J/m$^2$. Using Eq. \ref{DG}, this would predict a value of $\delta = 8.0$, in  good agreement with the one obtained from the fit of the experimental data, $\delta=10.1$.

{\it Discussion --}
This agreement supports the proposed picture of a shifted freezing transition, with wetting properties tuned by the electronic screening inside the confining substrates. There is  room for improvement of the theory, with a more complete description of the effect of the electronic screening on the RTIL wetting. This theoretical framework, which has not been developped up to now, will be the object of future work.

Our results also have implications for the question of dynamics of charging in dense electrolytes confined between metal surfaces, which is relevant to supercapacitors dynamics.  
Despite the importance of these phenomena for further developement of supercapacitors, there is so far a lack of experimental studies at the nanoscale, while unexpected phenomena were predicted at these scale \cite{Kornyshev2014b}. 
Our work underlines that nanoscale is a peculiar lengthscale for ionic liquid, and leads to strongly different behavior from what is observed at the macroscale. Our measurements unveil an overlooked phenomenon, suggesting that further improvements of the performances can be sought at the scales dominated by the atomic nature of matter.

In the context of lubrication, our results also suggest to take benefit of the dramatic and abrupt RTIL phase-change to tune nanoscale friction via modifications of the substrate, from insulating to metallic, and possibly with dedicated patterning of the metallic coating. The relatively weak solid phase indeed allows to avoid (undesired) direct substrate-substrate contact by generating strong normal forces. The solid phase can also be regenerated in situ, as it takes its origin in the RTIL confinement. 
Furthermore the modification of the confinement induced transition under voltage drop allows to modify finely the lubricating state by active 
polarization. While such perspectives require further exploration, they open new perspectives for phase-changing lubricants.

\noindent{\bf Acknowledgements} \\
L.B. and A.S. thank B. Rotenberg, B. Cross and E. Charlaix for many fruitful discussions. 
J.C., A.N. and A.S. acknowledge funding from the European Union's H2020 Framework Programme / ERC Starting Grant agreement number 637748 - NanoSOFT. 
L.B. acknowledges support from the European Union's FP7 Framework Programme / ERC Advanced Grant Micromegas. 
L.B. acknoweldges funding from a PSL chair of excellence.
Authors acknowledge funding from ANR project BlueEnergy.

\small

\end{document}